\documentclass{aa}
\usepackage{multirow}

\usepackage{multirow}
\usepackage{amsmath}
\usepackage{yhmath}
\usepackage{amssymb}
\usepackage{bm}
\usepackage[colorlinks,linkcolor=blue,anchorcolor=blue,citecolor=blue]{hyperref}
\usepackage{graphicx,subfig} 
\usepackage{lscape}
\usepackage{longtable}
\usepackage{txfonts}
\usepackage{natbib}
\usepackage{color}
\usepackage{booktabs}
\usepackage[dvipsnames,table,xcdraw]{xcolor}
\usepackage{comment}

\makeatletter
\renewcommand*\aa@pageof{, page \thepage{} of \pageref*{LastPage}}
\usepackage{hyperref}
\hypersetup{colorlinks=true,allcolors=[rgb]{0,0,0.8}}

\newcommand{\erosita}{eROSITA\xspace}

\newcommand{\esass}{\texttt{eSASS}\xspace}

\newcommand{\erass}[1][1]{eRASS1\xspace}
\newcommand{\extlike}{$\mathcal{L}_{ext}$\xspace}

\newcommand{\mbproj}{MBProj2D\xspace}

\usepackage[normalem]{ulem} 

\usepackage{natbib,twoopt}
\usepackage{xcolor}

\makeatletter
\newcommandtwoopt{\citeads}[3][][]{\href{http://adsabs.harvard.edu/abs/#3}%
    {\def\hyper@linkstart##1##2{}%
     \let\hyper@linkend\@empty\citealp[#1][#2]{#3}}}
  \newcommandtwoopt{\citepads}[3][][]{\href{http://adsabs.harvard.edu/abs/#3}%
    {\def\hyper@linkstart##1##2{}%
     \let\hyper@linkend\@empty\citep[#1][#2]{#3}}}
  \newcommandtwoopt{\citetads}[3][][]{\href{http://adsabs.harvard.edu/abs/#3}%
    {\def\hyper@linkstart##1##2{}%
     \let\hyper@linkend\@empty\citet[#1][#2]{#3}}}
  \newcommandtwoopt{\citeyearads}[3][][]%
    {\href{http://adsabs.harvard.edu/abs/#3}
    {\def\hyper@linkstart##1##2{}%
     \let\hyper@linkend\@empty\citeyear[#1][#2]{#3}}}
\makeatother
\usepackage{graphicx}
\usepackage{multirow}
\usepackage{siunitx}  
\usepackage{txfonts}
\title{The SRG/eROSITA All-Sky Survey}
\subtitle{Constraints on $f(R)$ Gravity from Cluster Abundance}
\titlerunning{Modified gravity with \erass}
\author{E. Artis\inst{1}, 
V. Ghirardini\inst{1}, 
E. Bulbul\inst{1}, 
S.~Grandis\inst{5,2},
C.~Garrel\inst{1},
N.~Clerc\inst{3},  
R.~Seppi\inst{1, 7},
J. Comparat\inst{1},
M. Cataneo\inst{4},
Y.~E.~Bahar\inst{1}, 
F.~Balzer\inst{1}, 
I.~Chiu\inst{8}, 
D.~Gruen\inst{5},
F.~Kleinebreil\inst{2, 4}, 
M.~Kluge\inst{1}, 
S.~Krippendorf\inst{5, 6}, 
X.~Li\inst{10},
A.~Liu\inst{1},
A.~Merloni\inst{1},
H.~Miyatake\inst{12, 13, 14},
S.~Miyazaki\inst{11},
K.~Nandra\inst{1},
N.~Okabe\inst{9}
F.~Pacaud\inst{4}, 
P.~Predehl\inst{1},
M.~E.~Ramos-Ceja\inst{1}, 
T.~H.~Reiprich\inst{4}, 
J.~S.~Sanders\inst{1}, 
T.~Schrabback\inst{2, 4}, 
S.~Zelmer\inst{1}, and
X.~Zhang\inst{1}
}
\institute{
Max Planck Institute for Extraterrestrial Physics, Giessenbachstrasse 1, 85748 Garching, Germany
\and
Universit\"at Innsbruck,  Institut f\"ur Astro- und Teilchenphysik, Technikerstr. 25/8, 6020 Innsbruck, Austria
\and
IRAP, Université de Toulouse, CNRS, UPS, CNES, F-31028 Toulouse, France
\and
Argelander-Institut f\"ur Astronomie (AIfA), Universit\"at Bonn, Auf dem H\"ugel 71, 53121 Bonn, Germany
\and
Universit\"ats-Sternwarte, LMU Munich, Scheinerstr. 1, 81679 M\"unchen, Germany
\and
Arnold Sommerfeld Center for Theoretical Physics, LMU Munich, Theresienstr. 37, 80333 M\"unchen, Germany
\and
Department of Astronomy, University of Geneva, Ch. d’Ecogia 16, CH-1290 Versoix, Switzerland
\and
Department of Physics, National Cheng Kung University, 70101 Tainan, Taiwan
\and
Department of Physical Science, Hiroshima University, 1-3-1 Kagamiyama,
Higashi-Hiroshima,Hiroshima 739-8526, Japan
\and 
McWilliams Center for Cosmology, Department of Physics, Carnegie Mellon University, Pittsburgh, PA 15213, USA
\and
Subaru Telescope, National Astronomical Observatory of Japan, 650 N Aohoku Place Hilo HI 96720 USA
\and
Kobayashi-Maskawa Institute for the Origin of Particles and the Universe (KMI), Nagoya University, Nagoya, 464-8602, Japan
\and
Institute for Advanced Research, Nagoya University, Nagoya 464-8601, Japan
\and
Kavli Institute for the Physics and Mathematics of the Universe (WPI), The University of Tokyo Institutes for Advanced Study (UTIAS), The University of Tokyo, Chiba 277-8583, Japan
}
\date{\today}

\titlerunning{Constraints $f(R)$ Gravity from \erass Cluster Counts}
\authorrunning{Artis et al.}

\begin{document}

\abstract{
The evolution of the cluster mass function traces the growth
of the linear density perturbations and can be utilized for
constraining the parameters of cosmological and alternative gravity models. 
In this context, we present new constraints on potential deviations from general relativity by investigating the Hu-Sawicki parametrization of the $f(R)$ gravity with the first SRG/\erosita\ All-Sky Survey (\erass) cluster catalog in the Western Galactic Hemisphere in combination with the overlapping Dark Energy Survey Year-3, KiloDegree Survey and Hyper Supreme Camera data for weak lensing mass calibration. For the first time, we present constraints obtained from cluster abundances only. When we consider massless neutrinos, we find a strict upper limit of $\log |{f_\mathrm{R0}}| < -4.31$ at 95\% confidence level. Massive neutrinos suppress structure growth at small scales, and thus have the opposite effect of $f(R)$ gravity. We consequently investigate the joint fit of the mass of the neutrinos with the modified gravity parameter. We obtain $\log |{f_\mathrm{R0}}| < -4.12$ jointly with $\sum m_\nu < 0.44\,\mathrm{eV}$ at 95\% confidence level, tighter than the limits in the literature utilizing cluster counts only. At $\log |{f_\mathrm{R0}}|= - 6$, the number of clusters is not significantly changed by the theory. 
 Consequently, we do not find any statistical deviation from general relativity from the study of eRASS1 cluster abundance.
Deeper surveys with eROSITA, increasing the number of detected clusters, will further improve constraints on $\log |f_\mathrm{R0}|$ and investigate alternative gravity theories.  
 }

\maketitle

\section{Introduction}
\label{sec:intro}

Our understanding of the Universe has been altered significantly with the discovery of its accelerated expansion, first demonstrated by supernovae observations \citep{Riess1998, Perlmutter1999}. The paradigm of the standard cosmological model widely represents this unknown dark energy with a constant added in Einstein's gravitational equations $\Lambda$, the so-called cosmological constant. Other ingredients necessary to describe our cosmological model include a non-relativistic (\textit{cold}) dark matter component that represents most of the mass fraction of the large-scale structures (LSS). These two components constitute the major fraction of the energy content of the present-day Universe and form the backbone of the concordance $\Lambda\mathrm{CDM}$ model. So far, when considered together with inflation, this parametrization is relatively successful at describing the main observed properties of the universe \citep{PlanckCollaboration2020}. 

Galaxy clusters represent the most massive virialized structures in the Universe. The number density of clusters of galaxies per unit mass and redshift, i.e., cluster mass function, and their spatial distribution, i.e., the two-point correlation function and higher order statistics, have great potential to constrain the nature of dark energy and test the robustness of the underlying cosmological models \citep[see][for a recent review]{Clerc2023}. Cluster counts are currently the most efficient at placing constraints on the matter density parameter ($\Omega_\mathrm{m}$), the root mean square of the density fluctuations in spheres of 8 $\textrm{Mpc} \ \textrm{h}^{-1}$ ($\sigma_\mathrm{8}$), and the sum of the masses of the neutrinos ($\sum m_\nu$). Additionally, it also breaks the degeneracies between the cosmological parameters of other probes, such as the baryon acoustic oscillations \citep[BAO;][]{Eisenstein2005, Alam2017}, redshift space distortions \citep[RSD;][]{Percival2009}, the weak-lensing cosmic shear \citep{Troxel2018, Asgari2021, Amon2022, Li2023}, Supernovae type Ia (SNe Ia) and other distance ladders \citep{Riess2019, Freedman2019}, and the cosmic microwave background \citep[CMB;][]{Bennett2013,PlanckCollaboration2020}. Previous and ongoing cluster surveys across a wide wavelength range demonstrate the potential of cluster number counts as a powerful cosmological probe when the biases related to mass calibration are reduced through the utilization of the weak gravitational lensing observations \citep[among others]{Mantz2015, Bocquet2019, Zubeldia2019, IderChitham2020, Garrel2022, Lesci2022, Sunayama2023, Fumagalli2023, Bocquet2023, Bocquet2024}. Modern cluster surveys, such as the \erosita All-Sky cluster survey \citep[see][]{Bulbul2024, Ghirardini2024}, are revolutionizing the field of cosmology, promising to reach the precision of the CMB experiments on cosmological parameters and have the potential to measure departures from Einstein's theory of general relativity (GR) and test theories of dark energy \citep{Wolf2023}. In addition to cluster abundance, \erosita is probing cosmology through the spatial distribution of galaxy clusters \citep{Seppi2024}, and the X-ray power spectrum \citep{Garrel2024}.
It also provides an avenue for testing alternative gravity theories, which could, in return, explain the late-time accelerated expansion of the Universe. 
Indeed, a way of explaining the Universe's accelerated expansion is to modify the theory of gravitation on large scales while retaining GR in the high-density solar system and galactic scales.   
Potential departures from GR can be modeled by a set of theories in which the Ricci scalar ($R$), the quantity describing the infinitesimal volumes in curved space-time, is replaced in the Einstein-Hilbert action (Equation~\ref{eq:EH-frgravity}) by a generic function $f(R)$. One of the models introducing a physically motivated function is the one proposed by \cite{Hu2007} (HS-$f(R)$ hereafter). This model has direct phenomenological implications. For example, \cite{Tsujikawa2009} and \cite{Gannouji2009} showed that it would significantly increase the structure formation rate, and thus, its predictions can be verified. In the context of cluster abundance studies, it increases the expected number density of massive dark matter halos (see figure \ref{fig:relative_difference_hmf}).
In order to observationaly constrain this model, many efforts have been made in recent years to provide tools computing summary statistics measuring departures from GR. First, numerous codes adapt the standard Boltzmann solvers, e.g., {\tt CAMB} \citep{Lewis2011CAMB} and {\tt CLASS} \citep{Lesgourgues2011} for modified gravity (MG hereafter) studies with alternative software such as {\tt EFTCAMB} \citep{Hu2014}, {\tt hi\_class} \citep{Zumalacarregui2017}, {\tt MGCAMB} \citep{Zucca2019, Wang2023}, or {\tt FRCAMB} \cite{Xu2015}. These extensions can compute the power spectrum in different MG scenarios. 

However, the hierarchical scenario of structure formation assumes that galaxy clusters originate from the gravitational collapse of matter overdensities. The cluster abundance can thus be obtained by computing the characteristics of this collapse in the corresponding theory \citep{Kopp2013, Lombriser2013} while considering the standard power spectrum. Starting from this fact, several studies attempt to constrain the $f(R)$ parameters using clusters in the literature. As the first of the studies, \cite{Schmidt2009} uses a sample of 49 low redshift clusters detected by ROSAT and re-observed by Chandra \citep{Vikhlinin2009} to constrain the HS-$f(R)$ model. This approach is combined with other cosmological probes by \cite{Lombriser2012}, showing that the best constraints are obtained when cluster abundance is considered. \cite{Cataneo2015} reported a robust upper limit for the value of the background scalar field $\log |f_\mathrm{R0}| < -4.79$ by utilizing a sample of 224 ROSAT-detected clusters combined with CMB data. Additionally, \cite{Lombriser2013} and later \cite{Cataneo2016} developed a halo mass function (HMF) formalism for predicting accurate cluster abundances, which was followed by \cite{vonBraun-Bates2017} and \cite{Gupta2022} formalisms. These studies explore the implication of $f(R)$ gravity in the halo formation process.
In parallel, \cite{Hagstotz2019} developed an alternative method where the HMF modeling accounts for the massive neutrinos and modified gravity effects. The modifications of gravity are then encapsulated in the collapse model and the halo mass function parametrization. In this work, we use this approach to predict the cluster counts. We emphasize again that all the cosmological quantities related to the matter power spectrum are computed with the standard version of {\tt CAMB}, without any need for an application of the modified Boltzmann solver.
With 5259 clusters of galaxies selected in its cosmology sample in a wide redshift range  $0.1 <z< 0.8$, the Western Galactic Half of the first eROSITA All-Sky Survey (\erass) has the statistical power to reach far beyond constraints available from any cluster survey in the literature. Consequently, this work utilizes the largest intracluster medium (ICM) selected cluster sample to date, in combination with weak lensing follow-up measurements for cluster masses and an accurate selection function to constrain deviations from GR by placing robust limits on the HS parametrization of $f(R)$ gravity.

The paper is organized as follows: in section \ref{sec:erass1}, we describe the datasets used. Section \ref{sec:mg_framwork} introduces the framework used to describe departures from GR in the HS-$f(R)$ gravity. Section \ref{sec:s_inference} provides a detailed description of the statistical methods. Finally, section \ref{sec:res} presents the constraints on the HS-$f(R)$ gravity and comparison with the $\Lambda\mathrm{CDM}$ concordance model. Throughout this paper, we use the notation $\log \equiv \log_\mathrm{10}$, and $\ln \equiv \log_\mathrm{e}$.

\section{Survey Data}
\label{sec:erass1}

The data used in this work is identical to the survey data presented in \citep{Bulbul2024, Kluge2024, Ghirardini2024, Grandis2024, Kleinebreil2023}. We summarize the multi-wavelength data, including X-ray, optical, and weak lensing observations, utilized in this work in the following section. 

\subsection{\erosita All-Sky Survey}

The soft X-ray telescope, \erosita, on board the Spectrum Roentgen Gamma Mission, completed its All-Sky Survey program on June 11, 2020, 184 days after the start of the survey. In this work, we use the data of the Western Galactic half of the \erosita\ All-Sky survey (359.9442~deg~$> l >$~179.9442~deg) belonging to the German \erosita\ consortium. \citet{Bulbul2024, Kluge2024} compiled two catalogs of clusters detected in this region: primary galaxy groups and clusters catalog and the cosmology sample based on the catalog of the X-ray sources in the soft band (0.2--2.3~keV) provided in \citet{Merloni2024}. We utilize the \erass\ cosmology sample in this work, which is compiled adopting a selection cut of extent likelihood, \extlike~$>6$ \citep[see][for further details]{Bulbul2024}. The optical identification of the clusters selected in the LS DR10-South common footprint produces a final cosmology sample of 5259 clusters of galaxies reaching purity levels of 95\%, an ideal sample for the modified gravity studies. Unlike the primary cluster catalog, the redshifts in the cosmology subsample are purely photometric measurements, allowing a consistent assessment of the systematics in our analysis. The total survey area covers a 12791~deg$^{2}$ region in the Western Galactic Hemisphere. The cont-rates are extracted with the 2D-image fitting tool \mbproj\ as described in \cite{Bulbul2024} and are employed as a mass proxy in the weak lensing mass calibration likelihood \citep{Ghirardini2024}.

\subsection{Weak Lensing Survey Data}
\label{subsec:weak_lensing_survey}

To perform our mass calibration in a minimally biased way, we utilize the deep and wide-area optical surveys for lensing measurements with the overlapping footprint with \erosita\ in the Western Galactic Hemisphere. The three wide-area surveys used in this work for mass calibration include the Dark Energy Survey (DES), Kilo Degree Survey (KiDS), and the Hyper SupremeCam Survey (HSC). We briefly describe the weak lensing survey data here.

The three-year weak-lensing data (S19A) covered by the HSC Subaru Strategic Program \citep{Aihara2018, Li2022} weak lensing measurements are made around the location of \erass clusters using in the $g$, $r$, $i$, $z$, and $Y$ bands in the optical. The total area coverage of HSC is $\approx500$~deg$^2$. The shear profile $g_{t}\left(\theta\right)$, the lensing covariance matrix that serves the measurement uncertainty, and the photometric redshift distribution of the selected source sample are obtained as the HSC weak-lensing data products for 96 \erass clusters, with a total signal-to-noise of 40.

We utilize data from the first three years of observations of the Dark Energy Survey (DES~Y3). The DES~Y3 shape catalog \citep{Gatti2021} is built from the $r, i, z$-bands using the \textsc{Metacalibration} pipeline \citep{Huff2017, Sheldon2017}. Considering the overlap between the DES~Y3 footprint and the \erass footprint, we produce tangential shear data for $2201$ \erass galaxy clusters, with a total signal-to-noise of 65 in the tangential shear profile. The details of the analysis and shear profile extraction are presented in \cite{Grandis2024}.

We use the gold sample of weak lensing and photometric redshift measurements from the fourth data release of the Kilo-Degree Survey \citep{Kuijken2019, Wright2020, Hildebrandt2021, Giblin2021}, hereafter referred to as KiDS-1000. We extract individual reduced tangential shear profiles for a total of 236 \erass galaxy clusters in both the KiDS-North field (101 clusters) and the KiDS-South field (136 clusters), as both have overlap with the \erass footprint with a total signal-to-noise of 19.

The weak lensing data in the mass calibration process is described in detail and published in \citep{Ghirardini2024, Grandis2024, Kleinebreil2023}. We do not process the data but adopt the framework from \citep{Ghirardini2024}.

\section{Modified Gravity Framework}
\label{sec:mg_framwork}

The evolution of the large-scale structures of the Universe is sensitive to modification to the standard frameworks of GR and $\Lambda\mathrm{CDM}$. To investigate potential discrepancies, we need to probe different scales and epochs. 
\begin{figure}
    \centering
    \includegraphics[scale=0.6]{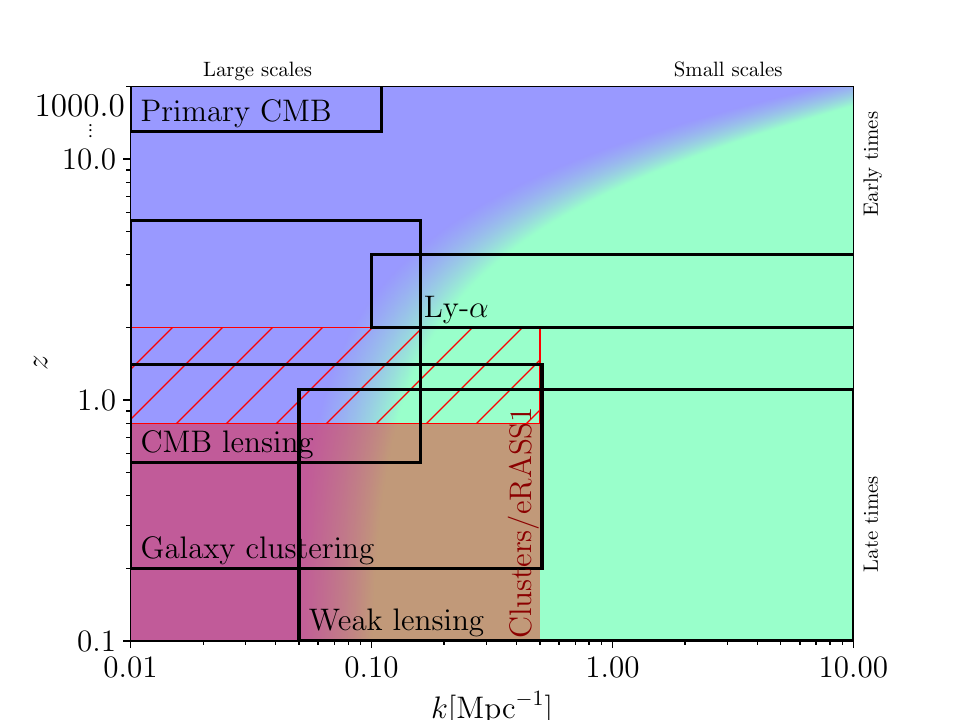}
    \caption{
    The estimation of the scale and redshift sensitivity of the different cosmological probes, adapted from \cite{Preston2023}. The background colors, from blue to green, represent the transition from the linear to the non-linear regime. We represent cluster abundance with the red rectangle. The cosmology sample of \erass spans the redshift range $0.1<z<0.8$, as represented by the solid red rectangle. Future data releases and experiments will increase the upper limit of this interval to the hatched red rectangle. Cluster abundance experiments thus probe a specific regime at large spatial scales and nearby Universe, offering a complementary window on structure formation. The experiments with cluster clustering \citep{Seppi2024} and X-ray power spectrum \citep{Garrel2024} are omitted here. In particular, the X-ray power spectrum might probe smaller scales.}
    \label{fig:cosmo_probes_scales}
\end{figure}
Figure \ref{fig:cosmo_probes_scales} shows the redshift and scale sensitivity of the most common cosmological probes. Galaxy cluster number counts efficiently place constraints on the low redshift/large-scale regime not explored by other probes. In this section, we describe the framework explored in this paper, namely the HS-$f(R)$ model.

\subsection{HS-$f(R)$ gravity in the context of cluster counts}
\label{subsec:HSfR_cc}
We choose to test the modified gravity model that consists of a modification of the Einstein-Hilbert action in the following form:
\begin{equation}
\label{eq:EH-frgravity}
S = \int \mathrm d x^4 \sqrt{-g} \left( \frac{R + f(R)}{16 \pi G} + \mathcal{L}_m \right) \: ,
\end{equation}
where $\mathcal{L}_m$ represents the Lagrangian of the matter field, $R$ is the Ricci scalar and $f(R)$ represents the extension to GR, for which $f(R) = -2\Lambda$. In the context of $f(R)$ gravity, many functional forms are proposed \citep[see][for a review]{DeFelice2010}. 
Throughout this paper, we focus on the \cite{Hu2007} model, which formalizes $f(R)$ as
\begin{equation}
\label{eq:HS_general}
    f(R) = -2\Lambda \frac{R^n}{R^n+m^{2n}},
\end{equation}
where $m^2$ represents the curvature scale, and $\Lambda$ and $n$ (the scaling index) are constants. We point out that the model does not include a cosmological constant as $f(R) \rightarrow 0$ when $R \rightarrow 0$. In the high curvature regime ($R \gg m^2$), equation~\eqref{eq:HS_general} becomes
\begin{equation*}
f(R) = -2\Lambda \frac{R^n}{R^n+m^{2n}} = -2\Lambda \frac{1}{1+\left(\frac{m^{2}}{R}\right)^n} \sim -2\Lambda \left(1-\left(\frac{m^{2}}{R}\right)^n\right).
\end{equation*}
To express this function in terms of $f_\mathrm{R}(R) =\mathrm{d}f/\mathrm{d}R$. Then, $f(R)$ becomes
\begin{equation*}
    f(R) = -2\Lambda - \frac{f_\mathrm{R0}}{n}\frac{\overline{R_0}^{n+1}}{R^n},
\end{equation*}
where we have defined value of the field $f_\mathrm{R0}= \mathrm{d}f/\mathrm{d}R \, (R_0) = -2n\Lambda m^{2n}/\overline{R_0}^{n+1}$ at z=0, $\bar R_0$ is the Ricci scalar at $z=0$ and overbars denote background quantities.
State-of-the-art numerical simulations are required to build a precise HMF suitable for cluster abundance cosmology. In these simulations, it is common to use $n=1$ \citep[e.g.][]{Oyaizu2008, Zhao2011, Giocoli2018}. For consistency with the HMF calibration method, we thus freeze the scaling index to unity.
We finally obtain
\begin{equation}
\label{eq:f_R}
f(R) \approx -2 \Lambda - f_{R0} \frac{\bar R_0^2}{R}.
\end{equation}
In this model, we note that when $|f_\mathrm{R0}| \rightarrow 0$, we recover the usual GR action with a cosmological constant in equation~\eqref{eq:EH-frgravity}. Thus, we interpret $\Lambda$ as the cosmological constant. The only parameter quantifying the deviations from GR is $f_\mathrm{R0}$. When $|f_\mathrm{R0}|\ll 1$, the background expansion is not distinguishable from GR. In this paper, our main goal is to measure $f_\mathrm{R0}$ or to set an upper limit on its value. In the rest of this section, we briefly describe the behavior of this model.

The variational principle applied on the action in Equation~\eqref{eq:EH-frgravity} leads to the modified Einstein's equation:
\begin{equation}
\label{eq:modified_einstein_eq}
G_{\mu \nu} - f_\mathrm{R} R_{\mu \nu} - \left( \frac{f}{2} - \Box f_\mathrm{R} \right) g_{\mu \nu} - \nabla_\mu \nabla_\nu f_\mathrm{R} = 8 \pi G T_{\mu \nu} \: . 
\end{equation}
As usual, the equation~\eqref{eq:modified_einstein_eq} provides the equation of motion of the scalar field $f_\mathrm{R}$:
\begin{equation}
\label{eq:field_fR}
\nabla^2 \delta f_\mathrm{R} = \frac{a^2}{3} \big( \delta R(f_\mathrm{R}) - 8 \pi G \delta \rho_m \big) \: ,
\end{equation}
 where $\delta$ denotes small perturbations of the related quantities. This also provides a Poisson-like equation given by
\begin{equation}
\label{eq:poisson}
\nabla^2 \psi = \frac{16 \pi G}{3} a^2 \rho_m - \frac{a^2}{6} \delta R(f_\mathrm{R}).
\end{equation}
For a better understanding of the behavior in this model, we provide the thin-shell condition example first introduced by \cite{Khoury2004} and adapted in \cite{Li2012} and \cite{Lombriser2013}. Since GR is well constrained in high-density environments like the solar system, we want to recover it at small scales: this is the so-called \textit{screening mechanism}. As an illustration, let us consider a top-hat overdensity of radius $R_\mathrm{in}$. Then, the gravitational pull exerted on a particle of mass $m$ outside of the overdensity can be expressed as
\begin{equation}
    \label{eq:grav_fr}
    F = G\frac{M_\mathrm{in}m}{r^2}\left(1+\frac{1}{3}\frac{\Delta R}{R_\mathrm{in}}\right),
\end{equation}
where $M_\mathrm{in}$ is the mass inside the top hat overdensity, $r$ is the distance to the center and $\Delta R / R_\mathrm{in}$ is the screening factor defined as
\begin{equation}
\label{eq:screen_fact}
\frac{\Delta R}{R_\mathrm{in}}=\mathrm{min}\left\{\frac{3|f_\mathrm{R}^\mathrm{in}- f_\mathrm{R}^\mathrm{out}|}{2\phi_\mathrm{N}} , 1\right\},
\end{equation}
where $\phi_\mathrm{N}$ is the Newtonian gravitational potential $\phi_\mathrm{N}=GM_\mathrm{in}/R_\mathrm{in}$, $f_\mathrm{R}^\mathrm{in}$ and $f_\mathrm{R}^\mathrm{out}$ are the values of the scalar filed inside and outside of the top-hat overdensity. Considering that the background expansion is considered in $\Lambda \mathrm{CDM}$, we can approximate the background Ricci scalar to the one expected in this model. Inside the top hat overdensity, we consider that the screening mechanism is active, i.e., we recover GR. We can then express the scalar field as 
\begin{equation}
    \label{eq:scal_value}
    f_\mathrm{R}^\mathrm{in} = f_\mathrm{R0}\left( \frac{1+4\Omega_\mathrm{\Lambda 0}/\Omega_\mathrm{m0}}{\rho_\mathrm{m}^\mathrm{ in}/\rho_\mathrm{m0}+4\Omega_\mathrm{\Lambda 0}/\Omega_\mathrm{m0}} \right),
\end{equation}
where $\rho_\mathrm{m}^\mathrm{in}$ is the density inside the considered region.
In equation~\ref{eq:scal_value}, we note that the scalar field values become smaller in high-density regions ($\rho_\mathrm{m0}\ll\rho_\mathrm{m}^\mathrm{in}$). Thus in Equation~\ref{eq:screen_fact}, the screening factor, becomes
\begin{equation*}
\frac{\Delta R}{R_\mathrm{in}}\sim\frac{3|f_\mathrm{R}^\mathrm{out}|}{2\phi_\mathrm{N}} .
\end{equation*}
The condition for the screening to be activated is $\Delta R / R_\mathrm{in} \ll 1$. Thus we must have that $|f_\mathrm{R}^\mathrm{out}|\ll \phi_\mathrm{N}$. Since $f_\mathrm{R}$ is a monotonic function of time, a relevant condition would be to have 
\begin{equation*}
    |f_\mathrm{R0}| < \phi_\mathrm{N}.
\end{equation*}
For massive galaxy clusters, the Newtonian potential (in natural units) is of order $10^{-5}$. Thus, for these gravitationaly bound systems, the deviation to GR is screened (i.e., suppressed) when $| f_\mathrm{R0}| < 10^{-5}$ (see section \ref{subsec:weak_mg}).
Overall, the HS-$f(R)$ gravity model is defined with only one additional parameter: $f_\mathrm{R0}$. This formalism allows the recovery of GR in the regimes necessary from observational constraints. In the next section, we describe the formalism predicting the dark matter halo abundance.

\subsection{The $f(R)$ halo mass function}
\label{subsec:fr_hmf}

The density of clusters per unit mass and redshift is modeled as follow:
\begin{equation}
    \label{eq:halo_mass_function}
    \frac{\mathrm{d}n}{\mathrm{d}\ln M} = \frac{\rho_\mathrm{m,0}}{M} \frac{\mathrm{d}\ln \sigma^{-1}}{\mathrm{d}\ln M}f(\sigma), 
\end{equation}
where $\rho_\mathrm{m,0}$ is the matter density at present, $f(\sigma)$ is the multiplicity function, and $\sigma(R,z)$ is defined by 
\begin{equation}
    \label{eq:sigma2_definition}
    \sigma^2(R, z) = \frac{1}{2 \pi^2}  \int_0^\infty k^2 P(k, z) \left( \frac{3 j_1(k R)}{k R} \right)^2 \,\mathrm{d}k ,
\end{equation}
where $j_1(x)=(\sin(x)-x\cos(x))/x^2$ is the spherical Bessel function of the first kind of order one, and $P(k)$ is the matter power spectrum.
We follow the approach developed in \cite{Hagstotz2019} to describe the specific behavior of the HMF in the HS-$f(R)$ framework when compared to the standard fitting functions and emulators \citep{Tinker2008, Despali2016, Bocquet2020, Castro2023} obtained from simulations assuming standard GR. The main advantage of this formalism is that the entire framework is derived from the collapse of the structures in $f(R)$ gravity, with standard computations for the power spectrum.
 Deviations from the GR HMF are modeled as follows: the multiplicity function can be written as in Equation~\eqref{eq:halo_mass_function}.
The goal is to correct the multiplicity function by a factor encapsulating the deviations from GR; therefore,

\begin{equation}
    \label{eq:correct_HMF}
    f^{f(R)}(\sigma) = f^\mathrm{GR}(\sigma) \cdot \eta(f_\mathrm{R0},\sigma),
\end{equation}
\noindent where the correcting factor is given by 
\begin{equation}
    \label{eq:correct_fact}
    \eta(f_\mathrm{R0},\sigma) = \frac{f_\mathrm{k}^{f(R)}(f_\mathrm{R0},\sigma)}{f_\mathrm{k}^\mathrm{GR}(\sigma)}.
\end{equation}
The the multiplicity functions including the subscript $k$ in the correcting factor are defined as follow.
In the formalism of the multiplicity function, the density field $\delta$ realizes a random walk (i.e., a Markov process) when smoothed by a simple top-hat filter in $k$-space, with the radius $R$ defined in Equation~\eqref{eq:sigma2_definition}. This means that in the trajectories $\delta(R)$, the next positions depend only on the current position. The multiplicity functions in Equation~\eqref{eq:correct_fact} are obtained that way and are noted with the subscript $k$. They are not realistic enough to describe the halo abundance. The $f(R)$ properties are encoded on the collapse of structures and not on the statistics of the density field, and their ratio is well suited to quantify departure from GR.
However, more realistic filters should be considered to derive a proper fitting function, which causes departures from the uncorrelated random walk. Any multiplicity function, including these non-Markovian corrections found in the literature, can be applied for $f^\mathrm{GR}$. In this work, we use the \cite{Tinker2008} as $f^\mathrm{GR}(\sigma)$. 
We emphasize that the resulting fitting function for HMF we use in this work was initially calibrated for masses defined as $M_\mathrm{200m}$. We convert the HMF to $M_\mathrm{500c}$ as \erass mass measurements are calibrated for $M_{500,c}$.
Additionally, doing so allows us to stay consistent with the standard $\Lambda\mathrm{CDM}$ cosmology analysis presented in \cite{Ghirardini2024}. 
In the case of GR, the multiplicity function is given by 

\begin{equation}
    \label{eq:non_markov_f}
   f_\mathrm{k}^\mathrm{GR}(\sigma) = \sqrt{\frac{2a}{\pi}}\frac{\delta_c}{\sigma}\exp\left( - a\frac{(\delta_\mathrm{c}+\beta\sigma^2)^2}{2\sigma^2} \right),
\end{equation}
\noindent, where $a$ and $\beta$ are fixed to the values provided in \cite{Hagstotz2019}, and $\delta_\mathrm{c}$ is the critical density for the collapse in GR given by the appendix C of \cite{Nakamura1997}

\begin{equation}
\label{eq:delta_c_GR}
\delta_c^\mathrm{GR}(z) = \frac{3 (12 \pi)^{2/3}}{20} \left(1 - 0.012299 \log \bigg( 1 + \frac{\Omega_m ^ {-1} - 1}{(1 + z)^3} \bigg) \right) \: .
\end{equation}
The multiplicity function in the case of HS-$f(R)$ gravity for non-markovian correction follows:

\begin{equation}
\label{eq:f_k_fR}
f^{f(R)}_k(\sigma) = \sqrt{\frac{2a}{\pi}} \frac{1}{\sigma} \mathrm{e}^{-a \bar B^2/(2 \sigma^2)} \left( \delta_\mathrm{c}^{f(R)} - \frac{3 M}{2} \frac{\partial \delta_\mathrm{c}^{f(R)}}{\partial M} \frac{\partial \ln \sigma}{\partial \ln R} \right) \: ,
\end{equation}

\noindent where $a$ is a constant. $\bar{B}$ is the barrier, the threshold beyond which collapse structures are formed. This threshold follows 
\begin{equation*}
    \bar{B} = \delta_\mathrm{c}^{f(R)}(f_\mathrm{R0},M,z) + \beta\sigma^2,
\end{equation*}
with $\beta$ being another constant.
The key of this formalism is thus to compute the spherical collapse threshold in $f(R)$ gravity. For a review, we point the reader to \cite{Kopp2013} and the previous references of this section. Here, we provide the numerical solutions used for the collapse model:

\begin{equation}
\label{eq:fR_barrier}
\delta_\mathrm{c}^{f(R)}(f_{R0}, M, z) = \delta_\mathrm{c}^\mathrm{GR}(z) \times \Delta \left(f_{R0}, M, z \right) \vphantom{\Big(} 
\end{equation}

\noindent where the $ \Delta \left(f_{R0}, M, z \right)$ is given by:

\begin{align}
    \label{eq:delta_fr_m_z}
    \Delta(f_{R0}, M, z) & = 1 + b_2 \left(1+z \right)^{-a_3} \left(m_\mathrm{b} - \sqrt{m_\mathrm{b}^2 + 1} \right) \\
     &+ b_3 \Big(\tanh\left(m_\mathrm{b} \right) - 1 \Big). \nonumber 
\end{align}

The different parameters of the collapse model are defined as follows, as they were originally fitted:

\begin{equation}
    \left\{    
     \begin{array}{ll}
         b_2 &= 0.0166\\
         a_3(f_{R0}) &= 1 + \exp \left( -2.08 \left(\log |f_{R0}| + 5.56 \right)^2 \right) \\
         m_b(f_{R0}, M, z) &= (1+z)^{a_3} \left( \log M - m_1(1+z)^{-a_4} \right) \\
        & m_1(f_{R0}) = \mu_1 \log |f_{R0}| + \mu_2 \\
        & a_4(f_{R0}) = \alpha_4 \Big( \tanh (0.69 \left(\log |f_{R0}| + 6.65 \right) ) \\
        & \quad\quad\quad\quad\quad +1 \Big)\\
         b_3(f_{R0}) &= \beta_3 \left(2.41 - \log |f_{R0}| \right)
    \end{array}
    \right.
\end{equation}

Three values are then required to describe the collapse threshold in the HS-$f(R)$ parametrization. Those are represented by the parameters $\mu_1, \mu_2, \alpha_4$ and $\beta_3$. In this work, we use the best-fit value given by \cite{Hagstotz2019}.

One can verify that this parametrization converges to the standard GR threshold for collapse $\delta_\mathrm{c}^\mathrm{GR}(z)$ \eqref{eq:delta_c_GR}, when $|f_\mathrm{R0}|$ takes small values.

The HMF derived here is computed for $M_\mathrm{200m}$, while the \erass cosmology pipeline considers $M_\mathrm{500c}$. Thus, we follow \citep{Hu2003} to convert the mass: we first compute the HMF from \cite{Tinker2008} $\mathrm{d}n/\mathrm{d}\ln M_\mathrm{500c}$ for a broad range of halo mass $M_\mathrm{500c}$
and convert these masses assuming the mass concentration relation from \cite{Duffy2008} and the Navarro, Frenk, and White (NFW) density profile \citep{Navarro1996}. The resultant $f(R)$ HMF thus follows
\begin{align}
    \label{eq:halo_mass_function_fr}
    \frac{\mathrm{d}n^{f(R)}}{\mathrm{d}\ln M_\mathrm{500c}} = & \frac{\rho_\mathrm{m,0}}{M_\mathrm{500c}}\left|\frac{\mathrm{d}\ln \sigma}{\mathrm{d}\ln M_\mathrm{500c}}\right|f_\mathrm{Tinker2008}(\sigma(M_\mathrm{500c})) \;  \nonumber \\
     & \times \frac{f_\mathrm{k}^{f(R)}(\sigma(M_\mathrm{200m}(M_\mathrm{500c})))}{f_\mathrm{k}^\mathrm{GR}(\sigma(M_\mathrm{200m}(M_\mathrm{500c})))}.
\end{align}
This equation represents the HMF used in this work, although we note that the uncertainties related to the HMF can impact our constraints \citep{ Salvati2020, Artis2021}.
\begin{figure}
    \includegraphics[scale=0.6]{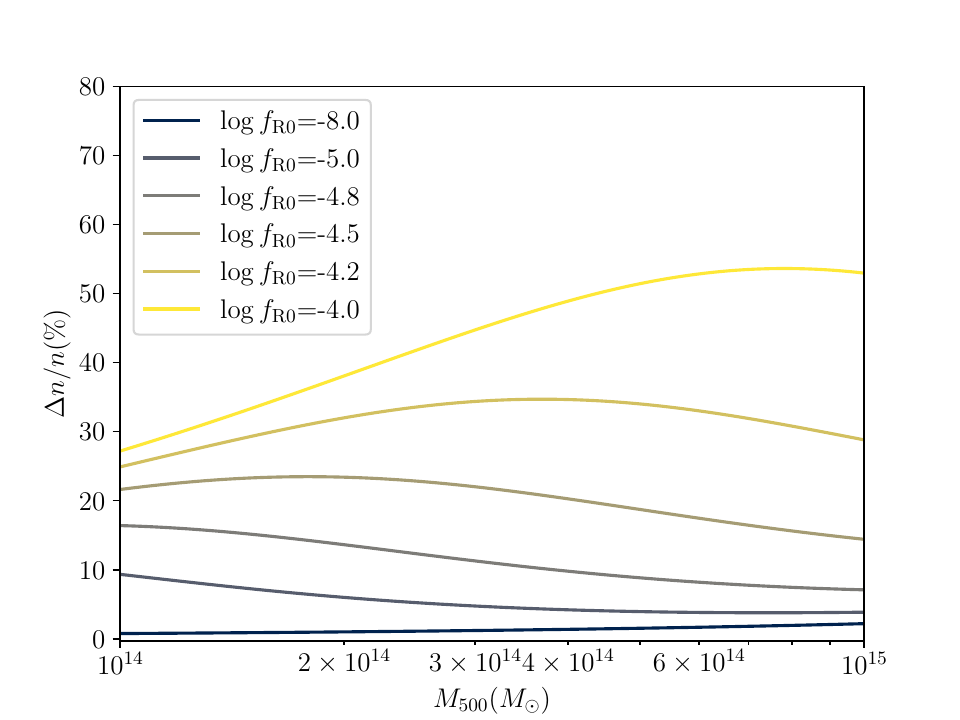}
    \caption{Relative difference (in percentage) of the HMF when $f_\mathrm{R0}$ increases. We note that $\Delta n = \mathrm{d}n/\mathrm{d}\ln M(f_\mathrm{R0})-\mathrm{d}n/\mathrm{d}\ln M$, where $\mathrm{d}n/\mathrm{d}\ln M(f_\mathrm{R0})$ is the HMF decribed with equation \ref{eq:halo_mass_function_fr} and $\mathrm{d}n/\mathrm{d}\ln M$ is obtained with the GR fitting function from \cite{Tinker2008}. Increased values of $f_\mathrm{R0}$ create more massive halos.}
    \label{fig:relative_difference_hmf}
\end{figure}
Figure~\ref{fig:relative_difference_hmf} shows the evolution of the HMF when $\log f_\mathrm{R0}$ increases. Due to different numerical simulations, halo finders, various fitting functions used in previous work, and baryonic effects, we expect the departure between the standard GR HMFs in the literature to be as high as 10\% \citep{Knebe2013, Castro2023}. Additionaly, the model used in this work presents up to 20\% differences with the predictions of \cite{Cataneo2016}. These systematic uncertainties are accounted for following  \cite{Costanzi2019} i.e. we compute the HMF as 

\begin{equation}
    \label{eq:corrected_hmf}
    \frac{\mathrm{d}\tilde{n}^{f(R)}}{\mathrm{d}\ln M_\mathrm{500c}} = \frac{\mathrm{d}n^{f(R)}}{\mathrm{d}\ln M_\mathrm{500c}} (s\ln (M/10^{14}M_\odot)+q),
\end{equation}

\noindent where $q$ and $s$ represent the errors respectively on the amplitude and slope of the HMF. Adopting a conservative approach, we use priors $q\sim\mathcal{N}(1,0.5)$ and $s\sim\mathcal{N}(0,0.5)$.

\subsection{Weak lensing mass calibration in modified gravity}
\label{subsec:weak_mg}

In this work, we aim to constrain modifications of the theory of gravity with the cluster halo mass function calibrated with the scaling relation between the X-ray count rate and the weak lensing shear data. Therefore, the predictions for the weak lensing signal of galaxy clusters in the modified gravity theories are needed for self-consistent analysis. As discussed in the following, we use the prediction from GR for the WL signal of massive clusters by simply utilizing the calibration performed by \citet{Grandis2024}. For readability, we provide the basis of the analysis and critical aspects \citep[see][for detailed description and review]{Bartelmann2001}.
In the framework of GR, the gravitational potential of a lens cluster deflects the light from distant background sources. The surface brightness $S_\mathrm{s}$ at position $\Vec{\theta}$ of a source with negligible dimensions compared to the lens size is as follows:
\begin{equation}
    \label{eq:suf_dist}
    S_\mathrm{obs}(\Vec\theta) = S_\mathrm{s}(\Vec{A}\Vec\theta),
\end{equation}
where $\Vec{A}$ is the lensing distortion matrix. It is written as 
\begin{equation}
\label{eq:dist_matrix}
\Vec{A} = 
\begin{pmatrix}
1 - \kappa - \gamma_1&  \gamma_2 \\
 \gamma_2 & 1 - \kappa + \gamma_1
\end{pmatrix} .
\end{equation}
The convergence $\kappa$ relates to isotropic transformations of the image, while the shear components $(\gamma_1, \gamma_2)$ are linked to the image distortion. The main transformations of the image are described by these two quantities, and they are computed as follow: the total distribution of mass in clusters is assumed to follow the NFW profile introduced in \cite{Navarro1996}, which is well fit- simulated DM, and defined as:

\begin{equation}
    \label{eq:NFW}
    \rho(r) = \frac{\displaystyle\delta(c_\Delta)\rho_\mathrm{c}(z)}{\displaystyle\frac{r}{r_\Delta/c_\Delta}\left(1+\frac{r}{r_\Delta/c_\Delta}\right)^2},
\end{equation}

\noindent where $r_\Delta$ is the radius defined such that the mean density of the enclosed volume is $\Delta$ times the mean density of the universe. $\rho_\mathrm{c}= 3H^2/8\pi G$ is the critical density and $c_\Delta$ is the concentration parameter. $\delta$ is a function that results from the definition of the radius $r_\Delta$ 
and reads:

\begin{equation}
    \label{eq:delta_function}
    \delta(c_\Delta) = \frac{\Delta}{3}\frac{c_\Delta^3}{\ln(1+c_\Delta) - c_\Delta(1+c_\Delta)}.
\end{equation}

This mass profile creates a gravitational potential noted $\phi$. We will note the lens cluster $l$ and background sources (typically galaxies). In the thin lens approximation (the typical size of the lens is negligible compared to the distance), we can write the lensing potential of a cluster as

\begin{equation}
    \label{eq:lensig potential}
    \psi(\Vec{\theta}) = \frac{2}{c^2}\frac{D_\mathrm{ls}}{D_\mathrm{l}D_\mathrm{s}}\int \phi (D_\mathrm{l}\Vec{\theta},z)\; \mathrm{d}z,
\end{equation}

\noindent where $\Vec{\theta}=(\theta_1, \theta_2)$ is the position on the sky. $D_\mathrm{ls},D_\mathrm{l},\mathrm{and}\; D_\mathrm{s}$ are respectively the distance between the lens cluster and the source galaxy, the distance between the observer and the lens, and the observer and the galaxy. 
The convergence and shear component can be expressed as a function of the derivatives of the lensing potential through

\begin{equation}
    \label{eq:der_lens_pot}
    \left\{
    \begin{array}{c}
         \kappa(\Vec{\theta})  = \displaystyle\frac{1}{2}\left( \frac{\partial^2 \psi}{\partial \theta^2_1} + \frac{\partial^2 \psi}{\partial \theta^2_1}\right)\\
         \gamma_1(\Vec{\theta}) =  \displaystyle\frac{1}{2}\left( \frac{\partial^2 \psi}{\partial \theta^2_1} - \frac{\partial^2 \psi}{\partial \theta^2_1}\right)\\
         \gamma_2(\Vec{\theta}) = \displaystyle \frac{\partial^2 \psi}{\partial\theta_1\partial\theta_2}
    \end{array}
    \right. .
\end{equation}

These expressions, combined with the definition of the lensing potential \eqref{eq:lensig potential}, lead to the following: if we define the surface mass density as 
\begin{equation}
\label{eq:surface_mass_density}
    \Sigma(R) = 2 \int_0^\infty \rho(R,z) \;\mathrm{d}z,
\end{equation}
where the matter density follows the NFW profile \eqref{eq:NFW}, the convergence profile is expressed as
\begin{equation}
    \label{eq:conv_equation}
    \kappa(R) = \frac{\Sigma(R)}{\Sigma_\mathrm{c}},
\end{equation}
and the radial dependence of the tangential shear follows
\begin{equation}
    \label{eq:shear_equation}
    \gamma_\text{t}(R) = \frac{\Sigma(<R)-\Sigma(R)}{\Sigma_\mathrm{c}},
\end{equation}
where $\Sigma(<R)$ is the mean surface mass density in the radius $R$, and the $\Sigma_\mathrm{c} = c^2/4\pi G \cdot D_\mathrm{s}/D_\mathrm{l}D_\mathrm{ls}$. These quantities are derived in \cite{Wright2000}. The weak lensing mass calibration is performed through the \textit{reduced tangential shear}
\begin{equation}
    \label{eq:reduced_shear}
    g_\mathrm{t}(R) = \frac{\gamma(R)}{1-\kappa(R)}.
\end{equation}
The reduced tangential shear profiles are related to the underlying mass through $P(\hat{g}_t | M_{\rm WL}, \hat{z})$ (see section \ref{subsec:weak_lensing_likelihood}). Since it is the consequence of the gravitational potential, it carries the necessary information on the underlying mass. To summarize, we assumed a standard lens equation resulting from GR to constrain our masses.

\noindent In this work, we consider one specific way of testing potential deviations: we probe the Hu-Sawicki parameterization of the $f(R)$ gravity. 

\noindent The theory of gravity impacts cluster WL in two ways: 
\begin{enumerate}
    \item the paths of light through the cluster's gravitational potential might be altered by the modified theory, leading to the same lensing potential \citep{Bekenstein1994} causing a different distortion in the shape of background galaxies and
    \item the effective Newtonian potential of galaxy clusters in a modified theory might deviate from the one predicted in GR.
\end{enumerate}
The first point is of no concern in the modification we explore. 
In the case of $f(R)$ gravity, the cosmic expansion remains the same (see section \ref{subsec:HSfR_cc}), but the geodesics are potentially affected. However, \cite{Schmidt2010b} showed that they remain virtually unchanged, up to a negligible factor of the order $\lesssim 10^{-4}$. We can thus use the generic lensing equations in HS-$f(R)$ gravity.

Halo matter profiles in $f(R)$ gravity have been extensively studied by \cite{Mitchell2018, Ruan23}, focusing on the mass concentration relation and the 3D density profiles. Neither of these two observables matches exactly the 2D projected matter density maps needed to anchor the WL mass calibration in a cluster cosmological context \citep{Grandis2021}. Nonetheless, the general findings of \citet{Mitchell2018} about the maximal halo mass at which halos become fully screened can still be applied. That work finds that halos with a mass larger than $10^{p_2}~h^{-1}\mathrm{M}_\odot$ are screened and thus behave like in GR. The parameter is given by
\begin{equation}
    p_2 = 1.50 \left( \log | f_\text{R0} | - \log(1+z) \right) + 21.6, 
\end{equation}
where we round before the numerical errors reported in \citet{Mitchell2018}. We shall evaluate this expression at $z=0.1$, the lower redshift limit of our cluster sample. Given that we are working with an X-ray-selected sample, this is also the redshift at which we reach the lowest limiting mass. For the values $\log_{10} | f_\text{R0} | = 5.25,\, 5.5,\, 6$ we find $p_2 = 13.69,\,13.31,\,12.56$. The first value of $p_2$ correspond to a mass of $10^{13.85}\mathrm{M}_\odot$. Considering that 95.4\% of the objects in the cosmology sample have an estimated mass greater than this value, we can argue that the cluster one halo regime that we analyze is also unaltered in the $f(R)$ gravity parameter space we explore.

\section{Statistical inference}
\label{sec:s_inference}

This section provides a brief description of the cluster cosmology inference pipeline. A detailed review can be found in the $\Lambda\mathrm{CDM}$ \erass cosmology results \citep{Ghirardini2024}. The main element is that the cluster abundance statistic follows a Poisson statistic (i.e., cluster abundance as a function of their observable is a Poisson process), coupled with weak lensing shear measurements, calibrating the scaling relation between mass and X-ray observable. A mixture model for eliminating the contamination due to active galactic nuclei (AGN) and background fluctuations misclassified as clusters in the eRASS1 cosmology subsample is employed to perform a precision cosmology experiment.

We express $\lambda$ as the intensity of the Poisson process (the number density of objects per unit of observable) and $x$ as the vector of the observables. This intensity depends on the model, and we note $\Theta$, the set of cosmological parameters. By definition, the expected number of objects whose observable properties belong to the sub-sample $\Omega$ of the observable parameter space follows;

\begin{equation}
    \label{eq:expected_number}
    N_{\{x\in\Omega\}}(\Theta) = \int_{\Omega} \lambda(x|\Theta)\;\mathrm{d}x.
\end{equation}

\noindent In our model, the global observable vector is defined as 

\begin{equation*}
    x = \{\widehat C_\mathrm{R}, \widehat z, \widehat\lambda, \mathcal{H}, g_+\},
\end{equation*}
where $\widehat C_\mathrm{R}$ are the observed count rate, $\widehat z$ are the observed photometric redshifts, $\widehat\lambda$ is the observed richness, $\mathcal{H}$ is the sky positions, which need to be considered due to the uneven exposure of the survey, and $g_+$ is the shear profile obtained from the different weak lensing surveys described in section \ref{subsec:weak_lensing_survey}.
The intensity of the process thus described the number density of the objects in the parameter space. The Poisson realization of the expected value $N_{\{x\in\Omega\}}(\Theta)$. In the framework that we adopt in this paper, we need to describe the intensity for the different species that we are considering to predict the abundance of the objects. Then, the general shape of the Poisson log-likelihood will follow; 

\begin{equation}
\label{eq:poisson_process}
\ln \mathcal{L}(\Theta) = \sum_i \ln (\lambda(x_i|\Theta))  - \int_x \lambda(x|\Theta)\, \mathrm{d}x,
\end{equation}

\noindent where the index $i$ runs over the observed objects. Using the Poisson likelihood, we neglect the cosmic variance \citep{Hu2003}  and the super sample variance \citep{Lacasa2019} as these effects are sub-dominant given the area covered by \erass and the limited number of clusters included in this analysis \citep[see][for discussion]{Ghirardini2024}. 
Our catalog contains three classes of objects: galaxy clusters (C), which are of interest, and contaminants, e.g., AGN, background fluctuations misclassified as clusters (NC). Our model simultaneously accounts for the cluster counts and the contaminant fractions through the Poisson mixture model. The total density is the sum of the three-component model;

\begin{equation}
    \label{eq:total_intensity}
    \lambda_\mathrm{tot}(x|\Theta) = \lambda_\mathrm{C}(x|\Theta) + \lambda_\mathrm{AGN}(x) + \lambda_\mathrm{NC}(x).
\end{equation}

In terms of the total number of objects in each class, we obtain

\begin{equation}
    \label{eq:def_mix_mod}
    N_\mathrm{tot}(\Theta) = N_\mathrm{C}(\Theta) + f_\mathrm{AGN}N_\mathrm{tot}(\Theta) + f_\mathrm{NC}N_\mathrm{tot}(\Theta),
\end{equation}
where $f_\mathrm{AGN}$ and $f_\mathrm{NC}$ are the respective fractions of contaminants. Consequently, we can express the number of AGN, false detections, and the total number of objects in the catalog as a function of the cosmology-dependent number of cluster $N_\mathrm{C}$:

\begin{equation}
    \left\{    
     \begin{array}{ll}
         N_\mathrm{tot}(\Theta) = (1/(1-f_\mathrm{AGN}-f_\mathrm{NC}))N_\mathrm{C}(\Theta) \\
         N_\mathrm{AGN}(\Theta) = (f_\mathrm{AGN}/(1-f_\mathrm{AGN}-f_\mathrm{C}))N_\mathrm{C}(\Theta) \\
         N_\mathrm{NC}(\Theta) = (f_\mathrm{NC}/(1-f_\mathrm{AGN}-f_\mathrm{NC}))N_\mathrm{C}(\Theta) \\ 
    \end{array}
    \right. .
    \label{eq:total_number}
\end{equation}

Starting from the formalism above, we can write the total number of objects is written as

\begin{equation}
    \label{eq:total_obs}
    N_\mathrm{tot}(\Theta) = \frac{1}{1-f_\mathrm{AGN}-f_\mathrm{NC}}\int_x \lambda_\mathrm{C}(x|\Theta) \;\mathrm{d}x,
\end{equation}
and that the number density of contaminants follows $\lambda_\mathrm{AGN}(x|\theta) = N_\mathrm{AGN}(\Theta)\mathcal{P}_\mathrm{AGN}(x)$, and $\lambda_\mathrm{NC}(x|\theta) = N_\mathrm{NC}(\Theta)\mathcal{P}_\mathrm{NC}(x)$. In this equation, $\mathcal{P}$ describes the probability distribution function of the respective object, depending on the observables. These terms are described in \ref{subsec:cont_fraction_likelihood}. Finally, the likelihood \ref{eq:poisson_process} becomes

\begin{equation}
    \label{eq:likelihood_cont}
    \begin{array}{ll}
    ln \mathcal{L}(\Theta) = & \displaystyle\sum_i \ln\; \Bigg(\lambda_\mathrm{C}(x_i|\Theta) \\
   &+ N_\mathrm{AGN}(\Theta)\mathcal{P}_\mathrm{AGN}(x_i) + N_\mathrm{NC}(\Theta)\mathcal{P}_\mathrm{NC}(x_i) \Bigg)  \\
    &- \,\displaystyle\frac{1}{1-f_\mathrm{AGN}-f_\mathrm{NC}}\int_x \lambda_\mathrm{C}(x|\Theta) \;\mathrm{d}x\; .
    \end{array}
\end{equation}
Separating the mass calibration likelihood from the number count likelihood is performed through the method described in the appendix of \cite{Bocquet2015}. For the clusters belonging to the footprints of the WL surveys, the number density becomes

\begin{equation}
    \label{eq:separ_lens}
    \lambda_\mathrm{C}(x_{\setminus\{g_+\}} ,g_+|\Theta) = \lambda_\mathrm{C}(x_{\setminus\{g_+\}}|\Theta)\mathcal{P}(g_+|x_{\setminus\{g_+\}},\Theta).
\end{equation}
This formalism is applied in Equation~\ref{eq:likelihood_cont} and allows for the calibration of X-ray and optical scaling relations. See section \ref{subsec:weak_lensing_likelihood} and \cite{Ghirardini2024} for the details of the mass calibration.

In the following sections, we briefly describe the components of Equation~\ref{eq:likelihood_cont}. The cluster abundance term $\lambda_\mathrm{C}$ is described in \ref{subsec:cluster_abundance_likelihood}, the weak lensing likelihood is given in section \ref{subsec:weak_lensing_likelihood}, and the mixture model accounting for contamination is provided in \ref{subsec:cont_fraction_likelihood}.

\subsection{Cluster abundance likelihood}
\label{subsec:cluster_abundance_likelihood}

The term $\lambda_\mathrm{C}$ of Equation~\ref{eq:likelihood_cont} represents the expected number density of clusters. It is obtained through
 \begin{equation}
\begin{aligned}
 \lambda_\mathrm{C}=\frac{\mathrm{d}N_\mathrm{C}}{\mathrm{d}\hat{C_\mathrm{R}}\mathrm{d}\hat{z}\mathrm{d}\hat{\mathcal{H}}} = & \int_z\int_M\int_{C_\mathrm{R}} \mathcal{P}(I| C_\mathrm{R}, z, \hat{\mathcal{H}}) \mathcal{P}(\hat{C_\mathrm{R}}|C_\mathrm{R})\\
   &\mathcal{P}(C_\mathrm{R}|M,z)\mathcal{P}(\hat{z}|z)\ \frac{\mathrm{d}n}{\mathrm{d}M} \frac{\mathrm{d}V}{\mathrm{d}z} \;\mathrm{d}M\mathrm{d}z\mathrm{d}C_\mathrm{R}.
\end{aligned}
\label{eq:xray_lik}
\end{equation}
$\mathrm{d}n/\mathrm{d}M$ is the halo mass function, $\mathrm{d}V/\mathrm{d}z$ is the comoving volume, and $\mathcal{P}(I| C_\mathrm{R}, z, \hat{\mathcal{H}})$ is the probability of detecting a cluster i.e. the selection function, the fraction of detected objects at a given observed sky position, redshift and count rates, given in \citep{Clerc2023b}. We note that the fraction of detected objects is estimated for a $\Lambda\mathrm{CDM}$ cosmology. We assume that this quantity is not significantly modified by the models considered in this work. The other terms are explained in the following.
Consistently with \citep{Ghirardini2024}, the count rates are assumed to be linked to the underlying mass of the objects through \citep{Ghirardini2024}

\begin{equation}
\label{eq:cr_mass}
\left<\ln \frac{\overline{C_\mathrm{R}}}{C_\mathrm{R,p}} \bigg| M, z\right> = 
\ln A_\mathrm{X} + 
b_\mathrm{X}(M, z) \cdot \ln \frac{M}{M_\mathrm{p}} + e_\mathrm{X}(z),
\end{equation}

\noindent where $C_\mathrm{R,p}=0.1\, \mathrm{cts}/\mathrm{s}$ and $M_\mathrm{p}  = 2\times 10^{14} M_\odot$ are the pivot count rates and mass. All the masses are expressed in solar masses $M_\odot$. The other terms are

\begin{equation}
\label{eq:bias_xray}
b_\mathrm{X}(M, z) = \bigg( B_\mathrm{X} + C_\mathrm{X} \cdot \ln \frac{M}{M_\mathrm{p}} + F_\mathrm{X} \cdot \ln \frac{1+z}{1+z_\mathrm{p}} \bigg),
\end{equation}

\noindent where $z_\mathrm{p}=0.35$ is the pivot redshift, and 

\begin{equation}
\label{eq:z_dependence}
e_\mathrm{X}(z) = D_\mathrm{X} \cdot \ln \frac{d_\mathrm{L}(z)}{d_\mathrm{L}(z_\mathrm{p})} + E_\mathrm{X} \cdot \ln \frac{E(z)}{E(z_\mathrm{p})} + G_\mathrm{X} \cdot \ln \frac{1+z}{1+z_\mathrm{p}},
\end{equation}

\noindent where $d_\mathrm{L}$ is the luminosity distance, $E(z) = H(z)/H_\mathrm{0}$ is the evolution function. The parameters $\{A_\mathrm{X}, B_\mathrm{X},C_\mathrm{X},D_\mathrm{X},E_\mathrm{X},F_\mathrm{X}\}$ are fitted jointly with the cosmological parameters to provide consistency and avoid astrophysical biases. 
We assume that the true count rates are the following, 
$$
C_\mathrm{R} \sim \mathrm{Log}\footnotesize{-}\mathcal{N}(\overline{C_\mathrm{R}(M,z)},\sigma_\mathrm{X}),
$$
 where $\mathrm{Log}\footnotesize{-}\mathcal{N}$ characterizes the log-normal distribution, $\overline{C_\mathrm{R}(M,z)}$, is the output of the scaling relation in Equation~\ref{eq:cr_mass}, and $\sigma_\mathrm{X}$ is the dispersion on the selection function, fitted together with the cosmological parameters. We also include the measurement errors on count rates fitted by the MultiBand Projector in 2D (\mbproj) tool \citep{Sanders2018}, and the observed count rates are 

$$
\hat{C_\mathrm{R}} \sim \mathrm{Log}\footnotesize{-}\mathcal{N}(C_\mathrm{R}, \hat{\sigma}),
$$

 \noindent where $\hat{\sigma}$ is fixed at the value given by the X-ray processing pipeline, MBProj2D \citep[see][for details]{Bulbul2024}. The observed redshifts are obtained through {\tt eROMaPPer}, the optical identification tool \citep[see][]{Kluge2024}  as in the following

 \begin{equation}
\begin{aligned}
P(\hat{z} | z) = & (1-c_z) \cdot \mathcal{N} (b_z \cdot z, \sigma_z (1+z)) + \\ & \quad + c_z 
 \cdot \mathcal{N} (b_z \cdot z + c_{\mathrm{shift}, z}, \sigma_z (1+z)),
\end{aligned}
\label{eq:zhat_given_z}
\end{equation}

 \noindent where $c_z, b_z, \sigma_z$ and $c_{\mathrm{shift}, z}$ are the fitted parameters of the distribution. 
The richness, proxy for the number of galaxies belonging to a given cluster, is computed starting from \cite{Grandis2021b} as

\begin{equation}
    \label{eq:richness}
    <\ln \lambda | M,z> = \ln A_\lambda + b_\lambda(z)\ln\left(\frac{M}{M_\mathrm{p}}\right) + C_\lambda \ln\left(\frac{1+z}{1+z_\mathrm{p}}\right),
\end{equation}
\noindent where all the masses are expressed in solar mass ($M_\odot$), and the redshift dependence of the mass slope follows

\begin{equation}
    \label{eq:richness_mass_slope}
    b_\lambda(z) = B_\lambda + D_\lambda\ln\left(\frac{1+z}{1+z_\mathrm{p}}\right).
\end{equation}

\noindent The optical richness is not one of the main observables used in the mass calibration. It is only used to eliminate the contamination in the sample through the mixture method. We keep the richness limit low ($\hat{\lambda}>3$) to ensure that no additional optical selection effects are introduced in the analysis. 

\subsection{Weak-lensing likelihood}
\label{subsec:weak_lensing_likelihood}

An essential step between the cluster number counts and X-ray observables obtained from the eROSITA survey and the halo mass function is the calibration of the cluster mass scaling relations. We utilize the weak lensing shear measurements in calibrating cluster scaling relations as it is currently the most reliable method with minimal bias available \citep{Giocoli2023}.
We use the surveys, i.e., the Dark Energy Survey (DES), the Kilo Degree Survey (KiDS), and the HyperSupreme Cam (HSC), presented in section \ref{sec:erass1} to achieve this goal. Following \citep{Grandis2024} we assume a relation between weak lensing inferred masses and true mass in the following form;

\begin{equation}
\left< \ln \frac{M_{\rm WL}}{M_\mathrm{p}} \bigg| M, z \right> = b(z) + b_{M} \ln \left( \frac{M}{M_\mathrm{p}} \right)
\label{eq:WLbias_given_M_z}
\end{equation}

The DES, KiDS, and HSC surveys are used simultaneously to calibrate the count rate to mass (Equation~\ref{eq:cr_mass}) and richness to mass (Equation~\ref{eq:richness}) relations. The scatter around this scaling is;
\begin{equation*}
\log \sigma_{M_{\rm WL}}^2 = s(z) + s_{M} \log \left( \frac{M}{M_\mathrm{p}} \right)
\end{equation*}

\noindent where the mass component is consistent with zero, and we fix $s_{M} = 0$ in the rest of the analysis as prescribed by  \cite{Grandis2024}. 

\subsection{Contamination fraction}
\label{subsec:cont_fraction_likelihood}

Although the eRASS1 cosmology subsample is a high-fidelity catalog with a purity level reaching 95\%, the remaining 5\% contaminants should be eliminated from our analysis. We apply a mixture model fitting for the fraction of AGN and random sources (RS). We refer the reader to \citep{Ghirardini2024} for a detailed method review.

\begin{table*}[h]
    \caption{Priors on the cosmological parameters used in this analysis when they are involved in the corresponding model. For the full set of priors used for the other parameters, including scaling relation and nuisance parameters, \citep[see][]{Ghirardini2024}}
    \label{tab:parameters_priors}
    \centering
    \begin{tabular}{llll}
    \hline
    \hline
    Parameter & Units & Description & Prior \\
    \hline
        $\Omega_{\mathrm{m}}$ & - & Mean matter density at present time & $\mathcal{U}(0.2, 0.95)$ \\
        $\log A_s$ & - & Amplitude of the primordial power spectrum & $\mathcal{U}(-10, -8)$ \\
        $H_0$ & $\frac{\frac{\rm km}{\rm s}}{\rm Mpc}$ & Hubble expansion rate at present time & $\mathcal{N}(67.77, 0.6)$ \\
        $\Omega_{\mathrm{b}}$ & - & Mean baryon density at present time & $\mathcal{U}(0.046, 0.052)$ \\
        $n_s$ & - & Spectra index of the primordial power spectrum & $\mathcal{U}(0.92, 1.0)$ \\
        $\sum m_\nu$ & eV & Summed neutrino masses &$\mathcal{U}(0, 1)$ \\
        $\log |f_\mathrm{R0}|$ & - & Logarithm of the derivative of $f(R)$ with respect to the Ricci scalar taken at present day & $\mathcal{U}(-8, -3)$ \\
    \hline
    \hline
    \end{tabular}
    \tablefoot{With $\mathcal{U}({\rm min}, {\rm max})$ we indicate a uniform distribution between `min' and `max'. With $\mathcal{N}(\mu, \sigma)$ we indicate a normal distribution centered on $\mu$ and with standard deviation $\sigma$.}
\end{table*}

\section{Results}
\label{sec:res}

In this section, we provide our results on the HS-$f(R)$ gravity based on cluster counts from the eRASS1 cosmology catalog with weak lensing mass calibration. We first show that our mass scale is consistent with the base $\Lambda\mathrm{CDM}$ analysis from \cite{Ghirardini2024}.

\subsection{Comparison of the mass scale with GR-$\Lambda\mathrm{CDM}$}

\begin{figure}
    \centering
    \includegraphics[scale=0.9]{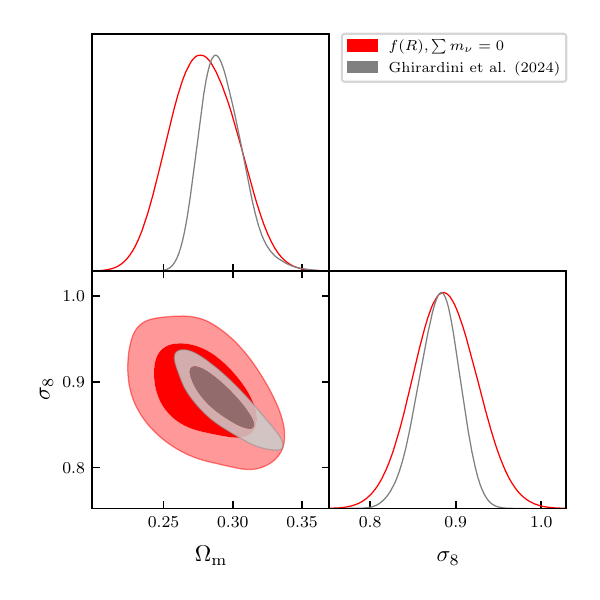}
    \caption{Comparison of the posteriors obtained on the $\Omega-\sigma_\mathrm{8}$ plane. We represent the $\Lambda\mathrm{CDM}$ analysis introduced in \cite{Ghirardini2024} in grey, while the posteriors of HS-$f(R)$ case are shown in red.}
    \label{fig:comp_om_s8_main}
\end{figure}

\begin{figure*}[!h]
    \centering
    \includegraphics[scale=0.9]{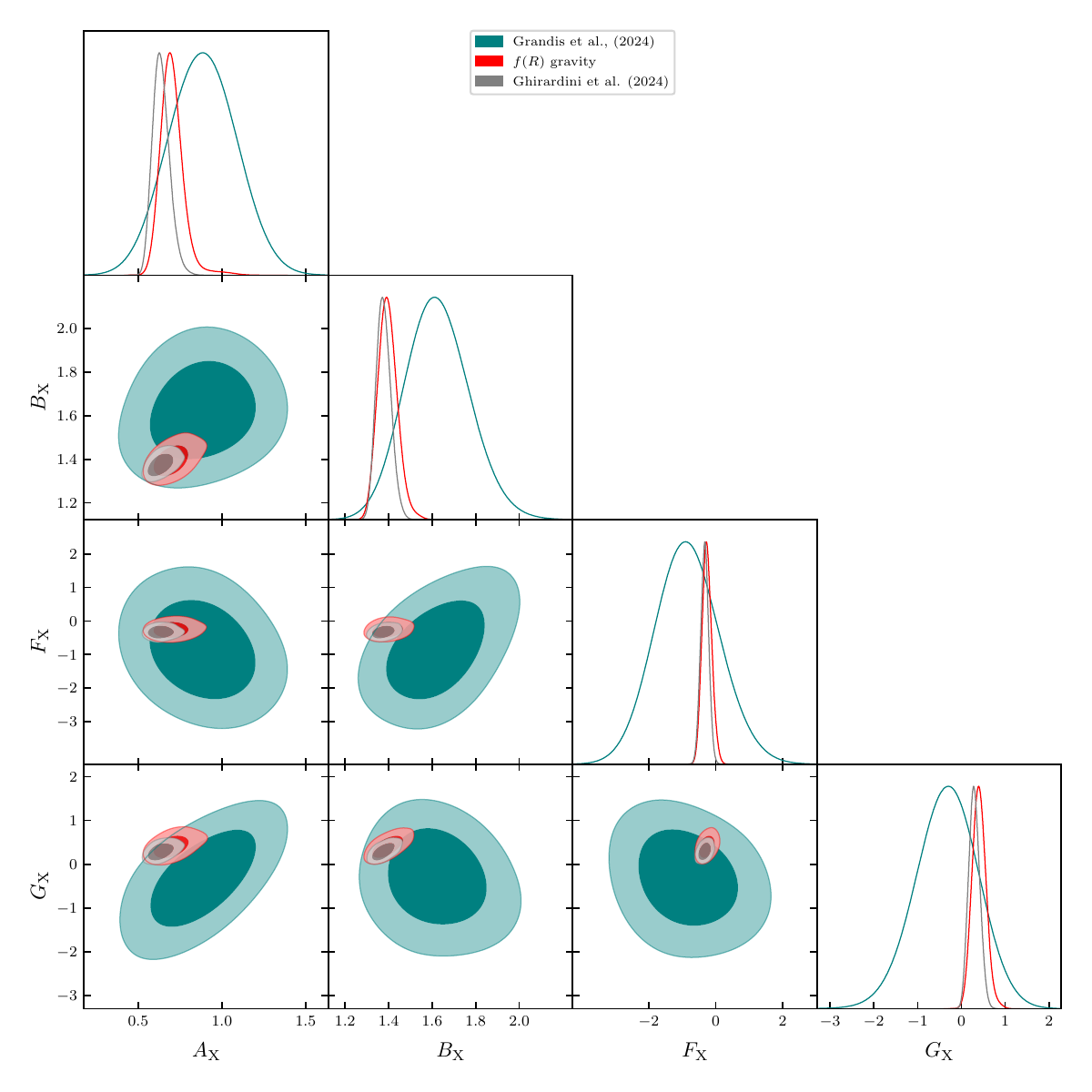}
    \caption{Comparison of the posteriors on the scaling relation obtained in the $f(R)$ framework (the mass of the neutrinos is fixed) and the standard $\Lambda\mathrm{CDM}$ analysis presented in \cite{Ghirardini2024}(grey). The different models are fully consistent with one another. We overplot the contours obtained from weak lensing only from \cite{Grandis2024}.}
    \label{fig:scaling_param}
\end{figure*}
One of the key ingredients of cluster abundance cosmology is the relation between the observable used to detect the objects and the underlying cluster mass. In this work, the main observable are the X-ray count rates of the clusters observed in the first \erosita All-Sky Survey. The modeling of the scaling relations between observable and cluster mass is described in section~\ref{sec:s_inference}. The parameters of this count rates to mass relation are fitted jointly with the cosmological parameters. This means that when considering extensions to general relativity that do not affect cluster masses due to screening mechanisms, we should find consistent scaling relations, as the fitted cosmology is similar to the one found in $\Lambda$CDM. 

The redshift dependence of the mass slope $C_\mathrm{X}$ is fixed to zero, the luminosity distance dependence $D_\mathrm{X}$ to -2, and normalized Hubble parameter dependence $E_\mathrm{X}$ to 2.
Figure ~\ref{fig:scaling_param} shows the fitted parameters $A_\mathrm{X}, B_\mathrm{X}, F_\mathrm{X}$ and $G_\mathrm{X}$ when compared to the standard $\Lambda\mathrm{CDM}$ analysis presented in \cite{Ghirardini2024} and the weak lensing mass calibration part of the likelihood presented in \cite{Grandis2024}. The latter represents the information coming only from the weak lensing mass calibration of DES clusters. 

Overall, there is good agreement between our different models whenever we consider deviation from GR. 
The mass calibration parameters are indeed not affecting our results.
This means that the mass calibration is robust and that changes in the cosmological model primarily affect the underlying distribution of clusters represented by the HMF.

\subsection{Constraints on $f(R)$ gravity}

\begin{table*}
\caption{Constraints on the HS-$f(R )$ gravity model in different scenarios. The errors provided for the constrained parameters are the 68\% confidence intervals. The upper limits are given at the 95\% confidence level.}
\label{table:HS_results}
\begin{center}
\begin{tabular}[width=0.5\textwidth]{c| c c c c c}
\hline\hline 
           & $\log |f_\mathrm{R0}|$ & $\Omega_m$ & $\sigma_\mathrm{8}$ & $\sum m_\nu$ & $S_\mathrm{8}$\\

\hline 
HS-$f(R)$  & $<-5.12$ & $0.28\pm 0.02$ &  $0.89\pm 0.04$ & -& $0.85\pm 0.04$\\
HS-$f(R)$ $+\sum m_\nu$  & $<-4.12$ &  $0.29\pm0.03$ &  $0.86\pm 0.05$ & $<0.39\,\mathrm{eV}$ & $0.83\pm 0.04$\\

\hline\hline
\end{tabular}
\end{center}
\end{table*}

\begin{figure}
    \centering
    \includegraphics[scale=0.6]{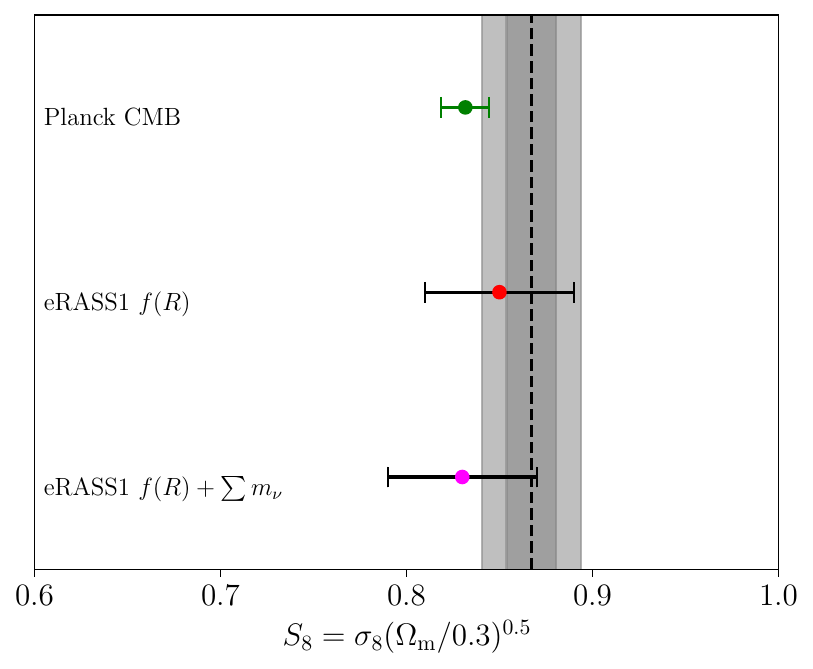}
    \caption{Comparison of the $S_\mathrm{8}$ parameter obtained using the different models. Wwe show the $1\sigma$ uncertainties. The case introducing massless neutrinos is in red, while the one considering massive neutrinos is shown in pink. Results from the cosmic microwave background are shown in green. The grey area represents the 68\% and 95\% confidence regions of the $\Lambda\mathrm{CDM}$ analysis from \cite{Ghirardini2024}.}
    \label{fig:s8_z}
\end{figure}

\begin{figure}
     \centering
     \includegraphics[scale=0.6]{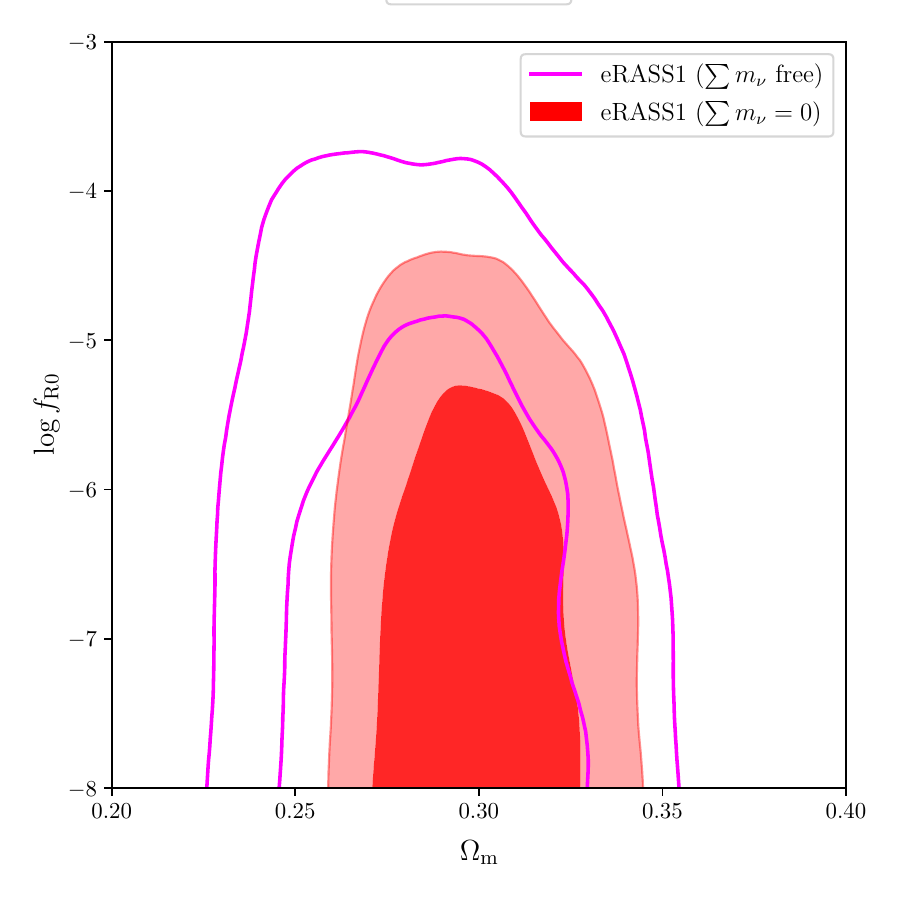}
     \caption{Constraints obtained on $\log f_\mathrm{R0}$ with \erass clusters. No external cosmological datasets are combined with our data shown in this figure. Contraints from with fixed neutrino mass are shown in red, while contraints where the neutrino mass is let free is shown in magenta.} 
     \label{fig:upper_fr0_lcdm}
 \end{figure}

We apply two different models to constrain the  HS-$f(R)$ gravity parameterization using the \citet{Hagstotz2019} formalism. In the first one, we assume massless neutrinos, while in the second one, we simultaneously constrain the sum of the neutrino masses and the parameter $\log |f_\mathrm{R0}|$ (Equation~\eqref{eq:f_R}). We summarize our results in Table~\ref{table:HS_results}. 
In both cases, \erass cluster number counts provide upper limits on $\log |f_\mathrm{R0}|$. In the first case, we find find the following constraints, summarized in figure \ref{fig:comp_om_s8_main}:
\begin{equation}
    \label{eq:constraints_om_s8_main_mu_fixed}
    \begin{array}{lcl}
     \Omega_\mathrm{m} &= & 0.28 \pm 0.02  \\
     \sigma_\mathrm{8} &= & 0.89 \pm 0.04 \\
     S_\mathrm{8} &= &  0.85 \pm 0.04
     
\end{array} ,
\end{equation}
 at 68\% confidence level. These constraints are in good agreement with the results of the standard $\Lambda\mathrm{CDM}$ analysis presented in \cite{Ghirardini2024}. Finally, we for the first time obtain constraints on the $f(R)$ parameters with cluster abundance only, obtaining:
\begin{equation}
    \label{eq:const_massless_neutrinos}
    \log |f_\mathrm{R0}| <-4.31 \; \mathrm{at}\; 95\%\;\mathrm{confidence}\; \left(\sum m_\nu =0\;\mathrm{eV} \right).
\end{equation}
Figure \ref{fig:upper_fr0_lcdm} presents the measurements in the $\Omega_\mathrm{m}-\log |f_\mathrm{R0}|$ plane. Our constraints show consistency with GR, as suggested by previous surveys using cluster abundance to probe HS-$f(R)$ gravity. When combined with CMB data, the most recent study by \cite{Cataneo2015} obtained $\log |f_\mathrm{R0}|<-4.79$. 
 Our results suggest that large unscreened departures from GR can be ruled out by cluster counts alone, thus providing an independent test of gravity on scales $\sim 10 \, {\rm Mpc}$.
Additionally, \cite{Hagstotz2019} showed that massive neutrinos can counteract the effect of modified gravity by reducing the abundance of massive structures. For instance, massive neutrinos increase the collapse barrier presented in equation \ref{eq:fR_barrier}. For $\sum m_\nu = 0.3\; \mathrm{eV}$, the collapse barrier is not distinguishable from the GR barrier if $\log |f_{R0}| = -5$. It is thus necessary to investigate the case where the sum of the mass of the neutrinos is fit jointly with $\log |f_\mathrm{R0}|$.
When we free the masses of the neutrinos, we obtain
\begin{equation}
    \label{eq:free_massive_neutrinos}
    \log |f_\mathrm{R0}| <-4.12 \; \mathrm{at}\; 95\%\;\mathrm{confidence}\; \left(\sum m_\nu \;\mathrm{free} \right),
\end{equation}
We also the current constraints on the cosmological parameters, including the neutrino masses:
\begin{equation}
    \label{eq:neutrino_mass}
    \begin{array}{lcl}
     \Omega_\mathrm{m} &= & 0.29 \pm 0.03  \\
     \sigma_\mathrm{8} &= & 0.86 \pm 0.05 \\
     S_\mathrm{8} &= &  0.83 \pm 0.04\\
     \sum m_\nu &< & 0.44 \;\mathrm{eV} \; \mathrm{at}\; 95\%\;\mathrm{confidence}\;
\end{array} ,
\end{equation}
These constraints are again fully consistent with those obtained in the base $\Lambda\mathrm{CDM}$ \citep{Ghirardini2024} (see table \ref{table:HS_results}).
We investigate the consistency between our $S_\mathrm{8}$ measurements and the one given in the standard $\Lambda$CDM analysis in \cite{Ghirardini2024}, who finds
\begin{equation}
    \label{eq:s8_ghirardini}
    S_\mathrm{8} = 0.86\pm 0.01 .
\end{equation}
Figure \ref{fig:s8_z} shows the different models, compared to the results obtained by \cite{PlanckCollaboration2020}. In all cases, our $S_\mathrm{8}$ value is in good agreement with the cosmic microwave background. Additionnaly, the three models are fully compatible.
However, the upper limits on the neutrino mass are higher due to the increased free parameters in the fits.
 It is worth noting that we solely use cluster abundances with no priors on the cosmological parameters when constraining $f_\mathrm{R0}$. The tighter upper limits reported in the literature, e.g., \cite{Kou2023}, find an upper limit of $\log |f_\mathrm{R0}| < -4.61$, combining the CMB measurements with galaxy clustering. 
The ability of finding upper limits with cluster counts alone can primarily be explained by the increased statistics provided by the \erass cluster sample. This allows us to provide a self consistent physical framework. Moreover, the constraints benefit from the shear measurement from high-quality DES, KIDS, and HSC mass calibration employed in this work \citep{Grandis2024, Kleinebreil2023}.
It is, therefore, clear that in both cases, whether the neutrino masses are allowed to be free, the constraints from the eRASS1 cluster count measurements remain consistent with GR with a cosmological constant, for which $f_\mathrm{R0} = 0$ as shown in equation \ref{eq:f_R}.


\section{Discussion and conclusions}
\label{sec:discussion}

In this work, we present results on potential deviations from standard GR and $\Lambda$CDM  using X-ray detected \erass clusters, in combination with LS DR10-South for optical confirmation and redshift measurement, and DES, KiDS, and HSC for weak lensing mass calibration. We derive constraints on the HS model of $f(R)$ gravity. We consider both the case of massive and massless neutrinos.

Overall, all the models that we have studied are statistically consistent with GR. 
Given the constraining power of \erass, we have chosen to present constraints from cluster abundance only. This choice allows us to measure our parameters with a consistent physical framework for the first time. Indeed, this works shows that cluster counts alone have a great potential to constraint modified gravity at the scale at which they are sensitive. Deeper surveys with eROSITA  will significantly increase the constraints that we obtain here and might rival the results obtained at smaller scales.
Overall, combined with the follow-up weak lensing observations, the first \erosita's All-Sky survey has the statistical power to constrain the cosmological parameters with percent level precision and test the general relativity at large scales.

Lastly, we investigate the consistency between our $S_\mathrm{8}$ measurements and the one given in the standard $\Lambda$CDM analysis and find good agreement. The modification of gravitation at the scale covered by cluster abundance does not significantly affect the $S_\mathrm{8}$ measurement. Our error bars are increased, but all posteriors are fully compatible. This emphasizes that the $S_\mathrm{8}$ measurement provided in the $\Lambda$CDM analysis is robust against modifications and in good agreement with \cite{PlanckCollaboration2020}. 

Finally, the precision of our analysis mostly relies on the halo mass function models available. We use the framework proposed by \cite{Hagstotz2019}, while others like \cite{Cataneo2016} propose alternative approaches. Additional work will be necessary to increase the precision of our constraints by increasing the reliability of the HMF, not only for $f(R)$ gravity, but also for any additional model of interest.

eROSITA, thus far, collected more than four All-Sky survey data. The final eRASS will detect about $10^5$ galaxy clusters and groups \citep[projected from the eFEDS results in][]{Liu2022,Bulbul2022}. The deeper eROSITA data and extensive cluster catalogs, in combination with state-of-the-art weak lensing mass calibration, will allow us to tighten our constraints on the $f(R)$ gravity models.

\begin{acknowledgement}

The authors thank Prof. Catherine Heymans for her help and availability for the implementation of the blinding strategy and her useful comments on the draft.

This work is based on data from eROSITA, the soft X-ray instrument aboard SRG, a joint Russian-German science mission supported by the Russian Space Agency (Roskosmos), in the interests of the Russian Academy of Sciences represented by its Space Research Institute (IKI), and the Deutsches Zentrum f{\"{u}}r Luft und Raumfahrt (DLR). The SRG spacecraft was built by Lavochkin Association (NPOL) and its subcontractors and is operated by NPOL with support from the Max Planck Institute for Extraterrestrial Physics (MPE).

The development and construction of the eROSITA X-ray instrument was led by MPE, with contributions from the Dr. Karl Remeis Observatory Bamberg \& ECAP (FAU Erlangen-Nuernberg), the University of Hamburg Observatory, the Leibniz Institute for Astrophysics Potsdam (AIP), and the Institute for Astronomy and Astrophysics of the University of T{\"{u}}bingen, with the support of DLR and the Max Planck Society. The Argelander Institute for Astronomy of the University of Bonn and the Ludwig Maximilians Universit{\"{a}}t Munich also participated in the science preparation for eROSITA.

The eROSITA data shown here were processed using the \esass software system developed by the German eROSITA consortium.
\\

V. Ghirardini, E. Bulbul, A. Liu, C. Garrel, S. Zelmer, and X. Zhang acknowledge financial support from the European Research Council (ERC) Consolidator Grant under the European Union’s Horizon 2020 research and innovation program (grant agreement CoG DarkQuest No 101002585). N. Clerc was financially supported by CNES. M. Cataneo acknowledges support from the Alexander von Humboldt Foundation and the German Centre for Cosmological Lensing. T. Schrabback and F. Kleinebreil acknowledge support from the German Federal
Ministry for Economic Affairs and Energy (BMWi) provided
through DLR under projects 50OR2002, 50OR2106, and 50OR2302, as well as the support provided by the Deutsche Forschungsgemeinschaft (DFG, German Research Foundation) under grant 415537506.
The Innsbruck group also acknowledges support
provided by the Austrian Research Promotion Agency (FFG)
and the Federal Ministry of the Republic of Austria for Climate Action, Environment, Energy, Mobility, Innovation and Technology (BMK) via the Austrian Space Applications Programme with grant numbers 899537 and 900565.

\\

The Legacy Surveys consist of three individual and complementary projects: the Dark Energy Camera Legacy Survey (DECaLS; Proposal ID \#2014B-0404; PIs: David Schlegel and Arjun Dey), the Beijing-Arizona Sky Survey (BASS; NOAO Prop. ID \#2015A-0801; PIs: Zhou Xu and Xiaohui Fan), and the Mayall z-band Legacy Survey (MzLS; Prop. ID \#2016A-0453; PI: Arjun Dey). DECaLS, BASS and MzLS together include data obtained, respectively, at the Blanco telescope, Cerro Tololo Inter-American Observatory, NSF’s NOIRLab; the Bok telescope, Steward Observatory, University of Arizona; and the Mayall telescope, Kitt Peak National Observatory, NOIRLab. Pipeline processing and analyses of the data were supported by NOIRLab and the Lawrence Berkeley National Laboratory (LBNL). The Legacy Surveys project is honored to be permitted to conduct astronomical research on Iolkam Du’ag (Kitt Peak), a mountain with particular significance to the Tohono O’odham Nation.

\\

Funding for the DES Projects has been provided by the U.S. Department of Energy, the U.S. National Science Foundation, the Ministry of Science and Education of Spain, the Science and Technology FacilitiesCouncil of the United Kingdom, the Higher Education Funding Council for England, the National Center for Supercomputing Applications at the University of Illinois at Urbana-Champaign, the Kavli Institute of Cosmological Physics at the University of Chicago, the Center for Cosmology and Astro-Particle Physics at the Ohio State University, the Mitchell Institute for Fundamental Physics and Astronomy at Texas A\&M University, Financiadora de Estudos e Projetos, Funda{\c c}{\~a}o Carlos Chagas Filho de Amparo {\`a} Pesquisa do Estado do Rio de Janeiro, Conselho Nacional de Desenvolvimento Cient{\'i}fico e Tecnol{\'o}gico and the Minist{\'e}rio da Ci{\^e}ncia, Tecnologia e Inova{\c c}{\~a}o, the Deutsche Forschungsgemeinschaft, and the Collaborating Institutions in the Dark Energy Survey.

The Collaborating Institutions are Argonne National Laboratory, the University of California at Santa Cruz, the University of Cambridge, Centro de Investigaciones Energ{\'e}ticas, Medioambientales y Tecnol{\'o}gicas-Madrid, the University of Chicago, University College London, the DES-Brazil Consortium, the University of Edinburgh, the Eidgen{\"o}ssische Technische Hochschule (ETH) Z{\"u}rich,  Fermi National Accelerator Laboratory, the University of Illinois at Urbana-Champaign, the Institut de Ci{\`e}ncies de l'Espai (IEEC/CSIC), the Institut de F{\'i}sica d'Altes Energies, Lawrence Berkeley National Laboratory, the Ludwig-Maximilians Universit{\"a}t M{\"u}nchen and the associated Excellence Cluster Universe, the University of Michigan, the National Optical Astronomy Observatory, the University of Nottingham, The Ohio State University, the OzDES Membership Consortium, the University of Pennsylvania, the University of Portsmouth, SLAC National Accelerator Laboratory, Stanford University, the University of Sussex, and Texas A\&M University.

Based on observations made with ESO Telescopes at the La Silla Paranal Observatory under programme IDs 177.A-3016, 177.A-3017, 177.A-3018 and 179.A-2004, and on data products produced by the KiDS consortium. The KiDS production team acknowledges support from: Deutsche Forschungsgemeinschaft, ERC, NOVA and NWO-M grants; Target; the University of Padova, and the University Federico II (Naples).

This paper makes use of software developed for the Large Synoptic Survey Telescope. We thank the LSST Project for making their code available as free software at  http://dm.lsst.org

\\

This work made use of the following Python software packages: 
SciPy\footnote{https://scipy.org/} \citep{Virtanen2020SciPy}, 
Matplotlib\footnote{https://matplotlib.org/} \citep{Hunter2007matplotlib}, 
Astropy\footnote{https://www.astropy.org/} \citep{Astropy2022}, 
NumPy\footnote{https://numpy.org/} \citep{Harris2020},
CAMB \citep{Lewis2011CAMB},
pyCCL\footnote{https://github.com/LSSTDESC/CCL} \citep{Chisari2019},
GPy\footnote{https://github.com/SheffieldML/GPy} \citep{gpy2014},
climin\footnote{https://github.com/BRML/climin} \citep{Bayer2015},
ultranest\footnote{https://github.com/JohannesBuchner/UltraNest/} \citep{Buchner2021}

\end{acknowledgement}

\bibliography{references.bib}
\end{document}